\documentclass[10pt,a4paper]{article}
\RequirePackage{amsmath,amssymb}
\RequirePackage[dvipsnames,usenames]{color}
\usepackage{cite}
\usepackage{fullpage}
\usepackage[british]{babel}
\usepackage[latin1]{inputenc}
\usepackage[T1]{fontenc}
\usepackage[final]{showkeys} 
\usepackage[bookmarks]{hyperref}
\usepackage{amsthm}
\usepackage{graphicx}
\usepackage{subfigure}
\usepackage{braket}
\usepackage{mathrsfs}
\usepackage{color}

\usepackage{bm}
\usepackage{amsfonts}
\usepackage{braket}

\newcommand{\be}{\begin{equation}}
\newcommand{\ee}{\end{equation}}
\newcommand{\e}{\epsilon}

\title{Unsettling physics in the quantum-corrected\\ Schwarzschild black hole}

\author{Valerio~Faraoni\thanks{E-mail: vfaraoni@ubishops.ca} ~and Andrea~Giusti\thanks{E-mail: agiusti@ubishops.ca}   
$\,$
\\
\\
{\em Department of Physics \& Astronomy, Bishop's University}
\\
{\em 2600 College Street, Sherbrooke Qu\'ebec, Canada J1M 1Z7}
}
\begin{document}
\def\theequation{\arabic{section}.\arabic{equation}} 

\maketitle

\begin{abstract}
We study a quantum-corrected Schwarzschild black hole 
proposed recently in Loop Quantum Gravity. Prompted by the fact that 
corrections to the innermost stable circular orbit of Schwarzschild 
diverge, we investigate timelike and null radial geodesics. Massive 
particles moving radially outwards are confined, while photons make it to 
infinity with infinite redshift. This unsettling physics, which deviates 
radically from both Schwarzschild (near the horizon) and Minkowski (at 
infinity) is due to repulsion by the negative quantum energy density 
that makes the quasilocal mass vanish as one approaches 
spatial infinity.
\end{abstract}

\newpage

\section{Introduction}
\label{sec:1}

Einstein's theory of gravity is plagued by singularities, which occur 
generically if the matter fields satify reasonable energy conditions 
\cite{Wald}. These singularities are not necessarily avoided even if 
quantum matter evades these energy conditions. However, it is believed 
that quantizing the geometry will cure the singularity problem. Loop 
Quantum Gravity and Loop Quantum Cosmology propose possible solutions in 
which quantum geometry avoids the classical singularities by making the 
curvature invariants finite and effectively discretizing spacetime. 
Referring specifically to black hole singularities,  the search for 
quantum-corrected black holes that are singularity-free has a long 
history, beginning with phenomenological attempts that were not solidly 
rooted in established theories, and later developing with full quantum 
gravity 
constructs. We refer the reader to the relevant literature, which is too 
long to summarize here. In broad strokes, one approach has been to 
quantize matter in curved space with the hope that this will be sufficient 
to avoid singularities, through the violation of the classical energy 
conditions.  However, whether this approach can be successful is now 
doubtful. Even if the energy conditions cherished by classical relativity 
are violated, there is no guarantee that singularities will be avoided.
 
The Loop Quantum Gravity approach is different, in that it attempts to 
quantize the geometry itself, and this more radical approach can, and 
should (at least in principle) be connected to the macroscopic world. For 
black holes, this statement means to connect to the physics in regions 
outside and around the horizons, as well as to spatial infinity far away 
from the black hole horizon.  In Loop Quantum Gravity, the spacetime 
geometry is 
fundamentally discrete, which is expressed by the fact that there is an 
area operator $\hat{\Delta} $ with a minimum eigenvalue that is stricly 
positive. All considerations about quantum black holes revolve around this 
basic fact and incorporate it somehow in the effective description of 
black holes. This discreteness of the geometry and the spectrum of the 
area operator must, therefore, leave some traces also in the macroscopic 
description of  a black hole at scales much larger than the Planck scale. 
Indeed, this is the case. Naturally, quantum-correcting the 
geometry of general relativity in the interior region ({\em i.e.}, 
inside 
the black hole horizon) leads to changes at the horizon and outside of it, 
when the interior geometry is matched to the exterior one (see, {\em 
e.g.}, \cite{HaywardBH, FrolovNotes, DeLorenzo:2014pta, DeLorenzo:2015taa, 
Giusti:2019wdx, Casadio:2017cdv,Casadio:2018qeh, Casadio:2019cux}). 
An unwanted 
effect is that, in some proposals, quantum-correcting the interior to cure 
the singularity may lead to large quantum effects in low curvature regions 
\cite{LQG1, LQG2, LQG3, LQG4, Suddho,Vagnozzi}. More in general, it is 
interesting to 
understand quantum corrections in the exterior regions which are, in 
principle, accessible to observers.

A quantum-corrected black hole which deviates from 
the Schwarzschild solution of general relativity was proposed recently in  
Refs.~\cite{AO1, AO2, AshtekarOlmedo}, with the explicit purpose 
of providing a singularity-free black hole which is also free 
of problems induced by large quantum corrections in the low-curvature 
regions. In this solution, which corrects the prototypical black 
hole, {\em i.e.}, the Schwarzschild geometry, loop 
corrections to the Schwarzschild geometry are quantified by a  small 
dimensionless parameter $\e$,  
which is mass-dependent. The explicit mass dependence is given by 
\cite{AshtekarOlmedo}
\be
\epsilon= \sqrt{1+\gamma^2 \delta_b^2}-1 \,,
\ee
where $\gamma \simeq 0.2375$ is the Barbero-Immirzi parameter (see 
\cite{AO1, AO2, AshtekarOlmedo}), 
\be
\delta_b = \left( \frac{\sqrt{\Delta}}{ \sqrt{2\pi} \, \gamma^2 m 
}\right)^{1/3}  \,,
\ee
$\Delta$ is the minimum positive eigenvalue of the area operator, given 
in terms of the Planck mass $l_{Pl}$ by $\Delta \simeq 5.17 \, l_{Pl}^2$, 
and 
$m$ is the black hole mass. Hence, when the macroscopic black hole 
properties are studied and one can take the limit $\delta_b \rightarrow 
0$, it is
\be
\epsilon \simeq \left( \frac{ \gamma^2 \Delta}{16\pi } \right)^{1/3} 
\frac{1}{m^{2/3}} \,.
\ee 
In the case of a solar mass black hole, 
this parameter is estimated to have values of order $\e \sim 10^{-26}$.  

To first order in the small parameter $\e$, the quantum-corrected black 
hole geometry of \cite{AO1, AO2, 
AshtekarOlmedo} is described by the the static and spherically symmetric 
line element\footnote{We follow the notations of Refs.~\cite{Wald, 
Carroll} and we use units in which Newton's constant $G$ and the speed of 
light are unity (but we occasionally restore $G$).}  (see 
Eqs.~(4.8)-(4.10) of Ref.~\cite{AshtekarOlmedo})
\be
ds^2=-\left( \frac{r}{r_S}\right)^{2\e}\left[ 1-
\left(\frac{r_S}{r}\right)^{1+\e}  \right] dt^2+
\frac{dr^2}{ 1-
 \left(\frac{r_S}{r}\right)^{1+\e}}  +r^2 d\Omega_{(2)}^2 \,,
\label{qbhmetric}
\ee
where $ d\Omega_{(2)}^2 =d\vartheta^2 +\sin^2 \vartheta \,  d\varphi^2$ is 
the 
line element on the unit 2-sphere, $r_S=2m$, and $m>0$ is a mass  
parameter analogous to the Schwarzschild mass. The line 
element~(\ref{qbhmetric}) reduces to the Schwarzschild one when $\e 
\rightarrow 0$. This happens even for small black holes 
\cite{AshtekarOlmedo}, but even more so when macroscopic black holes of, 
say, stellar mass are considered, due to the mass dependence $\epsilon 
\sim m^{-2/3}$.

Since $r$ is the areal radius, the possible horizons are located by 
the roots of $g^{rr} =\nabla^c 
r \nabla_c r=0$. There is a unique event horizon in the 
geometry~(\ref{qbhmetric}) and it coincides with the Schwarzschild horizon 
at  
$r_S=2m$. Contrary to Schwarzschild, there 
is an effective energy 
density outside the horizon, given 
by \cite{AshtekarOlmedo}
\be
\rho = - \frac{\e}{8\pi r^2} \left( \frac{r_S}{r} 
\right)^{1+\epsilon}\,,\label{quantumdensity}
\ee
which is negative and purely quantum-mechanical in origin.

In principle, correcting gravity due to geometry quantization or other 
reasons has implications for massive and massless particles and for fluids 
surrounding black holes and forming accretion disks around them. There is, 
therefore, much current interest in using observations of black holes to 
test deviations from general relativity or possibly detect scalar hair 
\cite{EHT,P1,P2,P3,P4,P5,Bertietal2013,sh2,sh3,sh4,sh5,sh6}.

Since particle motion near black hole horizons is relativistic and, 
in 
general, complicated, astrophysicists have introduced pseudo-Newtonian 
potentials to simplify the problem but still provide an effective 
description (at least for certain purposes) of timelike geodesics. 
Naturally, particles orbiting black holes in circular orbits are of 
special interest, and pseudo-Newtonian potentials can provide significant 
simplifications when studying their motions \cite{PaczynskiWiita, Kovar, 
Marek,hist1, hist2, hist3, hist4, hist5, hist6, hist7, hist8, hist9, 
hist10, hist11, hist12, TejedaRosswog14}. In spite of cheating many of the 
difficulties, pseudopotentials are still remarkably accurate in 
determining circular orbits. The phase space of massive test particles in 
the Schwarzschild geometry is very similar to that derived from the 
associated pseudo-potential \cite{phasespace}. Indeed, the 
pseudo-potential is precisely defined so that it preserves the 
equilibrium points of the relevant dynamical system \cite{PaczynskiWiita, 
Marek}. The Paczynski-Wiita potential \cite{PaczynskiWiita, Marek} was the 
first to be introduced in astrophysics and it locates exactly the 
innermost stable circular orbit (ISCO) and the marginally bound orbit of 
Schwarzschild, and it reproduces the Keplerian angular momentum $L(r)$. It 
is not as accurate in reproducing the Keplerian angular velocity and the 
radial epicyclic frequency, but it gives approximations that are 
nevertheless useful in some instances \cite{PaczynskiWiita, Marek}. In the 
following section we derive the pseudo-Newtonian potential associated with 
the quantum-corrected black hole~(\ref{qbhmetric}) and we show that 
the 
quantum correction to the Schwarzschild ISCO diverges. This fact prompts 
us to investigate radial timelike and null geodesics, to find that massive 
particles are confined by the negative quantum energy density and photons 
making it to infinity are infinitely redshifted. In Sec.~\ref{sec:3} 
we show that the Misner-Sharp-Hernandez/Hawking-Hayward quasilocal mass 
for this geometry vanishes as one approaches spatial infinity. We 
also report the Kodama quantities defined in spherical symmetry. 
Sec.~\ref{sec:4} contains the conclusions.

\section{Pseudo-Newtonian potential for the quantum-corrected black hole} 
\label{sec:2}

An analogue of the Paczynski-Wiita pseudo-Newtonian potential for the 
Schwarzschild black hole \cite{PaczynskiWiita, Marek} can be 
introduced for any static and spherically symmetric black hole  
\cite{ourpseudo}. This pseudo-Newtonian potential 
is \cite{ourpseudo}
\be
\Phi(r) = \frac{1}{2} \left( 1+\frac{1}{g_{00} } \right) \,,
\ee
which in our case becomes
\be
\Phi(r) = \frac{1}{2} \left[ 1-  \left(\frac{r_S}{r}\right)^{2\e} 
\frac{1}{ 1-  \left(\frac{r_S}{r}\right)^{1+\e}  } \right] \, ,
\ee
that approaches $1/2$ as $r \to \infty$ instead of vanishing.

Using the expansion  
\be
a ^{\alpha \e + \beta} = a^\beta \left[1 + \alpha 
\e \ln a + \mathcal{O}(\e ^2) \right] \, , 
\ee
with $a>0$ and $\alpha, \beta \in \mathbb{R}$,
one obtains, to first order, 
\begin{eqnarray}
\Phi (r) &=&  
- \frac{r_S}{2\left(r-r_S\right)} \left[
1 + \e \, \frac{r}{r_S}\, \frac{ (2r - r_S) }{(r - r_S)}  \, \ln 
\left( \frac{r_S}{r} \right) + \mathcal{O} (\e^2) 
\right] \\
&&\nonumber\\
&=& - \frac{m}{r-2m} \left[ 1+\e \, \frac{r}{m}  \frac{(r - m)}{(r - 
2m)}  \ln \left( 
\frac{2m}{r} \right)   \right] + \mathcal{O} (\e^2) \,. 
\label{pseudopotential}
\end{eqnarray}
As a check, one notes that for $\e \rightarrow 0$ this pseudopotential 
reduces to 
\be
\Phi_0 (r) = -\frac{r_S}{2r \left( 1-r_S/r \right) } =-\, \frac{m}{r-2m} 
\,,
\end{equation}
which is the well known Paczynski-Wiita pseudo-Newtonian potential for the 
Schwarzschild black hole \cite{PaczynskiWiita, Marek}. The 
$\epsilon$-expanded pseudopotential~(\ref{pseudopotential}) diverges as 
$r\rightarrow \infty$ instead of vanishing. Already at this stage 
one notices some problems with the asymptotics. The pseudo-Newtonian 
potential is defined also for non-asymptotically flat metrics ({\em e.g.}, 
(anti-)de Sitter and Schwarzschild-(anti-)de Sitter), and the fact that 
the 
$\epsilon$-expanded pseudopotential diverges signals problems with 
asymptotic flatness, which are discussed below.

The radii of the circular orbits in the metric~(\ref{qbhmetric}) are the 
roots of the 
equation \cite{ourpseudo, Marek}
\be
\frac{d \Phi}{dr}= \frac{L^2}{r^3} \,,
\ee
where $L$ is the angular momentum per unit mass of the particle on the  
circular orbit. 
This equation is what justifies the introduction of the ($\epsilon$-expanded)  
pseudo-Newtonian 
potential in the first place \cite{PaczynskiWiita, Marek}.  Using 
\be
\frac{d \Phi}{dr}= 
\frac{r_S}{2 \, (r-r_S)^2} + \frac{\e}{2 \, (r-r_S)^3} \left[
\left( 2 r - r_S \right) \left( r - r_S \right) + 
\left( 3 r - r_S \right) r_S \, \ln\left( \frac{r_S}{r} \right)
\right]  \,,
\ee 
the equation locating the circular orbits of massive test 
particles becomes
\be
\frac{r_S}{2 \, (r-r_S)^2} + \frac{\e}{2 \, (r-r_S)^3} \left[
\left( 2 r - r_S \right) \left( r - r_S \right) + 
\left( 3 r - r_S \right) r_S \, \ln\left( \frac{r_S}{r} \right)
\right] 
=
\frac{L^2}{r^3}    \,.\label{cazzo}
\ee
To zero order, these circular orbits satisfy
\be
\frac{r_S}{r_0}-  \frac{2L^2}{r_0^2} \left( 1-\frac{r_S}{r_0} \right)^2=0 
\,, \label{malumore}
\ee
while the radius  of a perturbed orbit is $r=r_0+\delta r_0$ with 
$\left|\delta r_0 /r_0 \right| =\mathcal{O}(\e)$.  Given the smallness of 
the parameter $\e$ quantifying the quantum gravity corrections to the 
Schwarzschild black hole, a linear expansion is an excellent 
approximation. By inserting the perturbed circular orbit radius 
$r_0+\delta r_0$, expanding eq.~(\ref{cazzo}) to first order, and 
taking advantage of the zero order equation~(\ref{malumore}), one 
obtains
\be
\frac{\delta r_0}{r_0} = \e \, r_0^3 \, \frac{\left( 2 r_0 - r_S \right) 
\left( r_0 - r_S \right) + \left( 3 r_0 - r_S \right) r_S \, \ln 
(r_S/r_0)}{2 \left[ 3L^2 (r_S - r_0)^3 + r_S \, r_0^4\right]}\,.
\ee  
Using again the zero order equation~(\ref{malumore}) to substitute for 
$L^2$ yields
\be
\frac{\delta r_0}{r_0} =
\e \, 
 \frac{\left( 2r_0 - r_S \right) \left( r_0 - r_S \right) +
r_S  \left(3r_0-r_S\right) \ln (r_S/r_0)}{
 r_S (3r_S - r_0 )} \, ;
\ee
the percent correction to the radii of the circular orbits is of first 
order in the parameter $\epsilon$. The ISCO of the Schwarzschild black 
hole lies at $r_0=6m$, for which the correction diverges.  This 
divergence shows that the $\epsilon$-correction has unwanted large effects 
no matter how small the parameter $\epsilon$, and is consistent with 
similar phenomenology found in Ref.~\cite{AndreaRoberto} for a different 
quantum-corrected black hole. 
To gain more insight, let us consider radial timelike and null geoesics.

Begin with a timelike radial geodesic followed by a particle of mass 
$m$, 4-velocity $u^c$, and 4-momentum $p^c=mu^c$: 
since $t^a \equiv \left( \partial/\partial t\right)^a $ is a timelike  
Killing vector, the energy $E$ of the particle is conserved along the 
geodesic, $ g_{ab} p^a t^b=-E$, yielding
\be
\frac{dt}{d\tau} =\frac{\bar{E}}{|g_{00}|} \,,
\ee
where $\bar{E} \equiv E/m$ is the particle energy per unit mass and  $\tau 
$ is the proper time along the geodesic. Substituting into 
the normalization $u_c u^c=-1$ gives 
\be
\left( \frac{dr}{d\tau} \right)^2 =\left( \frac{r_S}{r}\right)^{2\epsilon}
\left\{ \bar{E}^2  -\left( \frac{r}{r_S} \right)^{2\epsilon} \left[ 1- 
\left( \frac{r_S}{r}\right)^{1+\epsilon} \right] \right\} \,.
\ee
When the particle is at large distances from the horizon as $r\rightarrow 
+\infty$, one obtains
\be
\left( \frac{dr}{d\tau} \right)^2 \rightarrow -1 
\,,
\ee
which is absurd. The coordinate velocity of the particle is 
\be
v\equiv \frac{dr}{dt}=\frac{dr}{d\tau} \, \frac{d\tau}{dt} = 
\frac{dr}{d\tau}\, \frac{ \bar{E}}{|g_{00}|} \,,
\ee
which yields
\begin{eqnarray}
v^2 & \equiv & \left( \frac{dr}{dt} \right)^2 = \left( \frac{r}{r_S} 
\right)^{2\e} 
\left[ 1-\left( \frac{r_S}{r} \right)^{1+\e}\right]^2 
\left\{ 1-\frac{1}{\bar{E}^2} \left( \frac{r}{r_S} \right)^{2\e} \left[ 1- 
\left( \frac{r_S}{r} \right)^{1+\e} \right] \right\}  \nonumber\\
&&\nonumber\\
&\approx & -\frac{1}{\bar{E}^2} \left( \frac{r_S}{r} \right)^{4\e} \,;
\end{eqnarray}
also the coordinate velocity becomes  imaginary at sufficiently large 
radii, while it tends to zero as 
$r\rightarrow +\infty$.  The physical meaning is that the particle cannot 
be 
located at $r=\infty$ or at large radii. This surprising fact can 
be explained as follows: 
although decaying slightly faster than  $1/r^3$, the negative energy 
density~(\ref{quantumdensity}) repels a massive particle located at finite 
$r$ and prevents it 
from reaching infinity.  To make an analogy, consider the  
(uncorrected) Schwarzschild metric with 
negative mass parameter: a massive test particle will be repelled from a 
finite radius and go to infinity,                 
but in the present quantum-corrected geometry  the test 
particle is instead located at a finite radius and is repelled
by the effect of the negative quantum energy density far away. While the 
{\em local} effect of this negative energy density is negligible, the
{\em accumulated} effect of all the negative mass from this finite radius 
to infinity ``seen'' by the particle repels if {\em from infinity} and 
keeps it confined.

It is  now natural to ask what happens to an outgoing 
radial photon emitted at a  finite radius. Let $\lambda$ be an affine 
parameter along radial null geodesics. Then, energy conservation for these 
photons reads
\be
\frac{dt}{d\lambda} =\frac{E}{|g_{00}|} = \left( \frac{r_S}{r} 
\right)^{2\epsilon} \frac{E}{1-\left( r_S/r\right)^{1+\e} }  \,,
\ee
and substitution into the normalization $u_cu^c=0$ gives
\be
\left( \frac{dr}{d\lambda} \right)^2 =  E^2  \left( 
\frac{r_S}{r}\right)^{2\epsilon} \,.
\ee
As $r\rightarrow +\infty$, the photon slows down, $dr/d\lambda 
\rightarrow 0$  and gets ``tired'' ({\em i.e.}, infinitely redshifted), 
$dt/d\lambda \rightarrow 0$. This behaviour of test particles  shows 
that true asymptotic flatness is not achieved and there are 
lingering 
effects of the quantum corrections in the geometry. Indeed, 
these effects become more important as one goes further away from the 
horizon and ``sees'' more negative mass coming from the 
density~(\ref{quantumdensity}). 

Another effect induced by the negative energy density is that it will be 
impossible to introduce Painlev\'e-Gullstrand coordinates 
\cite{Painleve,Gullstrand} for the 
line element~(\ref{qbhmetric}). In fact, these coordinates are associated 
with observers starting at spatial infinity with zero velocity 
\cite{MartelPoisson} and they cannot be introduced in regions of negative 
quasilocal energy \cite{FaraoniVachon}.

\section{Kodama vector, Misner-Sharp-Hernandez mass, and Kodama 
temperature} 
\label{sec:3}

The quantum black hole metric~(\ref{qbhmetric}) is already written in the 
Abreu-Nielsen-Visser gauge 
\cite{NielsenVisser, AbreuVisser, mylastbook, ilnostroultimopaperdiobonino}, 
\be
ds^2=- \mbox{e}^{-2\Psi} \left(1-\frac{2M_\text{MSH}}{r} \right)dt^2 + 
\frac{dr^2}{1-2M_\text{MSH}/r } +r^2 d\Omega_{(2)}^2 
\ee
employing the areal radius as the radial coordinate. Here we have 
\be
\Psi= \e  \, \ln \left( \frac{r_S}{r} \right)\,;
\ee
this redshift function diverges as $r\rightarrow +\infty$ instead 
of going to zero as in Minkowski space.  $M_\text{MSH}$ is the 
Misner-Sharp-Hernandez mass of general relativity 
\cite{MSH1, MSH2}, which formally is always defined in a  spherically 
symmetric 
geometry  by
\be
1-\frac{2GM_\text{MSH}}{R}=\nabla^c R \nabla_c R =  g^{RR}  
\,,\label{somelabel}
\ee
where $R$ is the areal radius (which is a scalar and is well defined in 
spherical symmetry by using 
the area $A$ of the 2-spheres which are orbits of the rotational Killing 
vector field, 
$R=\sqrt{A}/(4\pi)$). (The last equality in~(\ref{somelabel}) holds 
in a coordinate system where $R$ is the radial coordinate, which is 
our case.) In spherical symmetry, 
the more 
general Hawking-Hayward 
quasilocal energy \cite{Hawking, Hayward}  reduces to the 
Misner-Sharp-Hernandez mass \cite{Haywardspherical}. 

In our case, the Misner-Sharp-Hernandez mass contained in a  sphere of 
radius $r$ is
\be
\label{FullMSH}
M_\text{MSH} (r) = \frac{r}{2}(1 - g^{rr}) = \frac{r}{2} 
\left(\frac{r_S}{r} \right)^{1+\e} \, ,
\ee
that, to the first order in $\e$, reduces to 
\begin{eqnarray}
M_\text{MSH} (r) &=&
\frac{r_S}{2} \left[ 1 + \e \, \ln \left(\frac{r_S}{r} \right) \right] + \mathcal{O} (\e^2) \nonumber\\
&&\nonumber\\
&=& m \left[ 1 + \e \, \ln \left(\frac{2 m}{r} \right) \right] + \mathcal{O} (\e^2) 
\,,\label{MSH}
\end{eqnarray}
where $m$ is the Misner-Sharp-Hernandez mass of the unperturbed general 
relativity (Schwarzschild) black hole, which 
coincides with the Schwarzschild mass. In regions near the 
horizon, the mass $M_\text{MSH}$ of the 
quantum-corrected black hole is smaller than that of the original 
Schwarzschild black hole, which is attributed to the negative 
quantum energy 
density~(\ref{quantumdensity}). Moreover, while the Misner-Sharp-Hernandez 
mass of Schwarzschild is the same ($m$) at any radius $r\geq r_S $, the 
quantum correction introduces a logaritmic dependence of $M_\text{MSH}$ on 
the radius at finite values of $r\geq r_S $. Besides, from 
Eq.~\eqref{FullMSH}
one finds that $M_\text{MSH}(r) \rightarrow 0$ as $r\rightarrow + \infty$. This 
prevents massive particles from reaching infinity and even tires photons, 
which reach infinity with zero frequency.

The relation between pseudo-Newtonian potential and Misner-Sharp-Hernandez 
mass is of some interest. 
The inversion of Eq.~(\ref{FullMSH}) gives
\be
m=\frac{r}{2}\left(\frac{2 M_\text{MSH}}{r} \right)^{\frac{1}{1+\e}} 
\ee
that, to first order in $\e$, yields
\be
m=M_\text{MSH} \left[ 1-\e \ln \left( \frac{2M_\text{MSH}}{r}\right) \right]
\ee
which, substituted into the expression~(\ref{pseudopotential}) of the 
($\e$-expanded) pseudopotential, yields
\be
\Phi(r) =-\,  \frac{M_\text{MSH}}{r-2M_\text{MSH}} \left[ 1 + \e \, 
\frac{r}{M_\text{MSH}}  \, 
\ln \left( \frac{2M_\text{MSH}}{r} \right) \right] \,.
\ee
For the Schwarzschild black hole, instead, one has \cite{ourpseudo}
\be
\Phi_{Schw}(r)= -\, \frac{ M_\text{MSH}}{ r-2M_\text{MSH} } \,; 
\label{PhiSchw}
\ee
 the difference 
arises because, for 
the quantum-corrected black hole,  $g_{00} \, g_{11} \neq -1 $ (cf. 
Ref.~\cite{ourpseudo}). Black holes that satisfy the condition 
$g_{00} \, g_{11}=-1$ (which has attracted some attention early on 
\cite{BondiKilmister60}), 
including the Schwarzschild geometry, have special 
geometric properties explored in \cite{Jacobson}. 
The condition $g_{00} \, g_{11}=-1$ characterizes spacetimes in which the 
(double) projection of  the 
Ricci tensor onto radial null vectors $l^a$ vanishes, 
$R_{ab}l^a l^b =0$ \cite{Jacobson}. An equivalent characterization is  
that the restriction of the Ricci tensor to the $\left( 
t,R\right)$ subspace is proportional to the restriction of 
the metric $g_{ab}$ to this subspace \cite{Jacobson}. 
A third characterization is that the areal radius  is an affine 
parameter along radial null geodesics \cite{Jacobson}. These 
characterizations are valid also  in higher-dimensional spacetimes.  
Solutions of the Einstein equations that satisfy this  
condition  include \cite{Jacobson} vacuum, electrovacuum with 
both Maxwell and  
non-linear Born-Infeld electrodynamics, and also a spherical 
global monopole called ``string hedgehog'' \cite{hedgehog1, hedgehog2}. 
Quantum-correcting the Schwarzschild black hole spoils all these 
geometric properties. This is an indication that $g_{00}\, g_{11}=-1$ is a 
rather fragile property and will likely be spoiled by all methods 
to quantum-correct classical metrics that satisfy it (notably, 
Schwarzschild), even if they are static 
like the geometry~(\ref{qbhmetric}) and have the same horizons of 
the uncorrected counterpart (which is not guaranteed in general 
since one can expect horizons to be dynamical or to fluctuate).

Let us consider now the Kodama quantities associated with spherical 
symmetry \cite{Kodama}.  The Kodama vector, always defined in spherical 
symmetry, is given by
\be
K^a = \mbox{e}^{\Psi} \left( \frac{\partial}{\partial t} \right)^a
\ee
in the Abreu-Nielsen-Visser gauge \cite{AbreuVisser, NielsenVisser, 
mylastbook}, therefore its components are
\be
K^{\mu} = \Big(  \left(\frac{r_S}{r}\right)^\e ,0,0,0 \Big)
\ee
and it is parallel, but not equal, to the timelike Killing vector. The 
Kodama four-current in this gauge is \cite{AbreuVisser, NielsenVisser, 
mylastbook}
\be
J^{\mu} = \frac{2\, \mbox{e}^{\Psi}}{r^2} \Big( -\frac{d M_\text{MSH}}{dr} 
,  \frac{d M_\text{MSH}}{dt} , 0,0 \Big) =
 \Big( \frac{\e}{r^2} \left(\frac{r_S}{r}\right)^{2\e+1} 
,0,0,0 \Big) \simeq 
\e \, \Big( \frac{r_S}{r^3} ,0,0,0 \Big) \,.
\ee
The Kodama temperature at the horizon $r=r_S=2m$ is proportional to the 
Kodama surface gravity $\kappa_\text{Kodama}$:
\begin{eqnarray}
T_\text{Kodama}& =& \frac{ \kappa_\text{Kodama} }{ 2\pi} \Big|_H= 
\frac{ \mbox{e}^{-\Psi (r_H)}}{2\pi} \left[
\frac{1-2M_\text{MSH}' (r_H)}{2r_H}\right]\nonumber\\
&&\nonumber\\
& =& \frac{1}{2\pi} \left( \frac{ 1 + \e}{2 \, r_S} \right) 
\nonumber\\
&&\nonumber\\
&=& \frac{1}{8\pi m} \left(1 + \epsilon \right) \,.
\end{eqnarray}
Since the metric is static, the Kodama temperature coincides with the 
Killing temperature 
found in \cite{AshtekarOlmedo} using Euclidean methods 
associated with periodicity in Euclidean time.

\section{Conclusions}
\label{sec:4}

We have examined the pseudo-Newtonian potential for the black hole 
of Refs.~\cite{AO1,AO2,AshtekarOlmedo} derived in Loop Quantum Gravity.  
The simple relation~(\ref{PhiSchw}) between pseudo-Newtonian potential and 
this quasilocal mass becomes complicated (albeit only to first order in 
the quantum corrections) and makes the pseudopotential stabilize to $1/2$ as 
$r\rightarrow +\infty$. The deviation from the corresponding 
Schwarzschild ISCO is dramatic:  to first order in $\e$,  
the percent correction to the radius of this orbit diverges, which signals 
severe deviations from Einstein's theory. Prompted by this 
fact, we have examined radial timelike and null geodesics, finding that 
massive test particles in radial motion starting from finite radii never 
make it to infinity, while outgoing photons arrive to infinity with 
infinite redshift. We 
trace these effects to the repulsion of the negative quantum energy 
density~(\ref{quantumdensity}). Unlike in Schwarzschild spacetime, the 
Misner-Sharp-Hernandez/Hawking-Hayward 
quasilocal mass is position-dependent and vanishes as $r\rightarrow +\infty$. 

Quantum-correcting the Schwarzschild black holes changes its asymptotics 
and makes it lose some of its peculiar features in the region outside the 
horizon. One should recover with an excellent approximation general 
relativity in the strong gravity region and Minkowskian physics as 
$r\rightarrow +\infty$, but this is not the case.  In other words, even  
extremely small quantum corrections alter radically the physics of 
Minkowski space far away from the horizon. The problems discussed in previous literature with 
quantum-correcting black holes \cite{LQG1, LQG2, LQG3, LQG4} 
seems to persist in  
this Loop Quantum Gravity black hole, even though  it was supposed to be free 
from 
such problems. 

Another problem, although not as important, consists of understanding why 
the physics deviates so strongly from the Schwarzschild physics near the 
horizon and from the Minkowski one at spatial infinity. The answer 
possibly lies in the fact that a proper limit of spacetimes as a parameter 
(in our case, $\e$) varies should not be based on coordinates, but should 
be done in an invariant way. Long ago, Geroch warned that the limit of the 
Schwarzschild geometry as the mass diverges is either the Minkowski space 
or a Kasner space \cite{Geroch}. A coordinate-independent approach based 
on the Cartan scalars has been developed in general relativity \cite{26} 
and then applied to the limit of Brans-Dicke gravity when the Brans-Dicke 
parameter $\omega$ diverges \cite{27}. This aspect will be explored in 
future work.

\section*{Acknowledgments}
This work is supported by the Natural 
Sciences and Engineering Research Council of Canada (Grant No. 2016-03803 
to V.F.) and by Bishop's University. The work of A.G. has been carried out in 
the framework of the activities of the Italian National Group for Mathematical
Physics (GNFM, INdAM).

\end{document}